# Possible Sources of Iron Nuclei in Ultra-High-Energy Cosmic Rays

**A. V. Uryson***

*Lebedev Physical Institute, Russian Academy of Sciences, Moscow, 117997 Russia*

**Abstract**—The origin of a cosmic-ray particle with an ultra-high energy of $2.44 \times 10^{20}$ eV recorded by the Telescope Array is discussed by assuming that it is a heavy nucleus, namely an iron or silicon nucleus. Iron is chosen as the heaviest element in cosmic rays, while silicon is chosen to find out how much the theoretical results obtained vary with the mass of the heavy nucleus. The constraints on the distances to the possible extragalactic sources of a heavy particle with such an energy are analyzed by assuming an extragalactic origin of this particle and using a routine calculation of the origin of cosmic rays in intergalactic space. These sources are discussed.

*E-mail: uryson@sci.lebedev.ru

## INTRODUCTION

In this paper we discuss the origin of a cosmic-ray particle with an ultra-high energy of $2.44 \times 10^{20}$ eV recorded by the Telescope Array (Telescope Array Collaboration 2023) assuming that this particle is a heavy nucleus, namely an iron or silicon nucleus. Iron is chosen as the heaviest element in cosmic rays, while silicon is chosen to find out how much the theoretical results obtained vary with the mass of the heavy nucleus.

The sources of ultra-high energy (UHE) cosmic rays (CRs) with $E > 10^{19}$ eV have not yet been firmly established. Both galactic and extragalactic objects, where CR particles can be accelerated to such energies, are discussed in the literature.

In this paper we assume that CRs are of extragalactic origin. Among the possible extragalactic sources of UHECRs, for example, gamma-ray bursts, starburst galaxies, galaxy clusters, radio galaxies, BL Lac objects, and supermassive black holes (SMBHs) are discussed in the literature.

We assume that CRs are accelerated to UHEs in the vicinity of SMBHs. CRs can be accelerated in jets (Biermann 1997; Istomin and Gunya 2020), in accretion disks (Haswell et al. 1992), and near the polar caps of SMBHs, where the particles fall from the accretion disk (Kardashev 1995; Neronov et al. 2009); CR particles can also be accelerated in the ergosphere of SMBHs (Tursunov et al. 2020; Comisso and Asenjo 2021). The SMBH jet and accretion disk contain stellar matter and, therefore, elements with various mass numbers $A$ up to iron nuclei are probably present in CRs in their sources.

The GZK-effect for nuclei manifests itself as UHECR nuclei propagate in intergalactic space: the elemental composition of CRs changes, since the CR particles interact with background radiations — the cosmic microwave background, radio emission, and extragalactic light. This involves the following reactions:

$$A + \gamma \rightarrow A + e^{+} + e^{-}, \quad (1)$$

(direct pair production, the threshold center-of-mass energy is of 1MeV)

$$A + \gamma \rightarrow A^{'} + mN + n\pi, \quad (2)$$

(photopion production,the threshold center-of-mass energy is of 145 MeV)

$$A + \gamma \rightarrow A^{'} + mN, \quad (3)$$

(the photodisintegration of nuclei, the threshold energy is tens of MeV)

The photodisintegration of UHE nuclei in intergalactic space was analyzed theoretically by Stecker and Salamon (1999).

The elemental composition of UHECRs is investigated by indirect methods using ground-based arrays: the mass number of a CR particle is determined from the depth of the maximum in the number of particles in the extensive air shower produced by it. According to the results obtained by the Telescope Array (Abbasi et al. 2024), the Pierre Auger Observatory (Abdul Halim et al. 2025), and the Yakutsk array (Glushkov et al. 2024), heavy nuclei, including iron nuclei, are present in UHECRs. As has already been said, the most energetic particle with an energy of 2.44 $\times$ $10^{20}$ eV was recorded by the Telescope Array, and this particle is probably a heavy nucleus (Telescope Array Collaboration 2023).

In this paper we discuss the possible sources of particles with such an energy by assuming that they are heavy nuclei. We consider iron and silicon in UHECRs, which mass numbers are $A$ = 56 and 28, respectively. The silicon nuclei in UHECRs can be the primary ones, i.e., the nuclei emitted by sources, and the secondary ones, i.e., those produced in intergalactic space by nuclei heavier than silicon, in reactions (2) and (3).

In this paper we consider, first, the primary silicon nuclei and, second, the silicon nuclei that were produced in reactions (2) and (3) by primary iron nuclei in intergalactic space.

We analyzed the primary iron and silicon nuclei that reached the Earth in the following way. As a result of reactions (1)–(3), not all primary nuclei, but only some of them and their fragments reach the array. We calculated the fraction $R$ of particles with the mass number $A$ = 56 that reached the array from their possible source relative to all of the remaining particles with mass numbers from $A$ = 1 to 56:

$R$ = (particles with $A$ = 56) / (all particles with $A$ = 1−56). (4)

We analyzed the secondary silicon nuclei produced by iron nuclei in a similar way by calculating

$P$ = (secondary particles with $A$ = 28) / (all particles with $A$ = 1−56). (5)

Our calculations were performed for various path lengths of the particle from the source and, thus, we obtained constraints on the distances to the possible sources based on which they are discussed.

In our calculations we used the TransportCR code (Kalashev and Kido 2015) publicly accessible on the internet, in which the photonuclear interactions are calculated with their cross sections from Puget et al. (1976).

THE MODEL

The main assumptions of the model concern the sources of UHECRs and the interaction of particles with background radiations.

Various models of possible UHECR sources and background radiation parameters have been investigated in the literature. For our calculations we chose the model of sources from Kachelries et al. (2017), since it describes the set of data on extragalactic CRs. Let us describe our model.

The parameters of the CR sources were the following.

(1) UHECRs are accelerated in point extragalactic sources.

(2) The injection spectrum in the sources is a power-law one, $E^{-\gamma}$, $\gamma = 2.2$. The maximum particle energy in the source is $10^{21}$ eV. We estimated the minimal particle energy $E_{\rm min}$ in the source as follows. A particle with charge $Z$ will not be confined by the magnetic field of the source galaxy and will be able to leave it if the Larmor radius $R_B$ of the particle in the galaxy's magnetic field $B$ exceeds the approximate thickness of the galactic disk $h_d$, $R_B \geq h_d$ (see, e.g., Ginzburg and Syrovatskii 1963):

$$E_{\rm min} = 300 R_B Z B, \quad (6)$$

where the energy $E_{\rm min}$ is measured in eV, the Larmor radius $R_B$ is in cm, and the magnetic field $B$ is in G.

Assuming that the sizes and magnetic fields of the most source galaxies are approximately the same as those of our Galaxy, i.e., the disk thickness is $h_d \approx 230$ pc and the magnetic field is $B \approx 10^{-6}$ G, we obtain the following minimum energies of the iron and silicon nuclei that escape from the source galaxy: $E_{\rm min\ Fe} \approx 5.6 \times 10^{18}$ eV for iron ($Z$ = 26) and $E_{\rm min\ Si} \approx 3.0 \times 10^{18}$ eV for silicon ($Z$ = 14).

(3) In our calculations we took into account the evolution of the sources. The evolution of SMBHs is unclear, and in our calculations we took the evolution of one of the types of active galactic nuclei — BL Lac.

(4) UHECRs are iron ($A$ = 56) or silicon ($A$ = 28) nuclei. Of course, both protons and nuclei of other elements are present in UHECRs. However, we discuss the possible sources of UHE iron and silicon particles and, therefore, consider the propagation of only these particles in intergalactic space.

The model assumptions about the background radiations were taken to be the following.

(1) The cosmic microwave background (CMB) has a Planck energy distribution, the mean photon energy is $\varepsilon_r = 2.3 \times 10^{-4}$ eV, and the photon density is $n_r = 400$ cm$^{-3}$.

(2) The background radio emission has the characteristics derived by Protheroe and Biermann (1996, 1997).

(3) The extragalactic background light (EBL) has the characteristics given by Inoue et al. (2013).

## RESULTS

The energy dependence $R(E)$ for primary iron and silicon nuclei near the Earth for several source redshifts $z$ is shown in Fig. 1. We will discuss the dependence $R(E)$ for nuclei with energies $E \geq 10^{20}$ eV. In this range the value of $R$ decreases with increasing particle energy.

We are interested in the energy range of nuclei in which $R \gtrsim 0.01$. As $z$ increases, this range decreases and is shifted toward lower energies. Clearly, the value of $R$ also decreases with increasing distance to the source. The energy ranges of iron nuclei in which $R \gtrsim 0.01$ and the value of $R$ for several source redshifts are given in Table 1; the analogous results for silicon nuclei are given in Table 2. It follows from Table 1 that iron nuclei with an energy of $2.44 \times 10^{20}$ eV and $10^{20} \leq E < 2.44 \times 10^{20}$ eV can arrive at the array from a distance $L \approx 1$ and $\approx$40 Mpc, respectively. Silicon nuclei with an energy of $2.44 \times 10^{20}$ eV can also arrive at the array from a distance $L \approx 1$ Mpc.

The energy of the particle being discussed is $E$ = 244 EeV and was determined with a statistical error of ±29 EeV and a systematic error of (+51, −76) EeV (Telescope Array Collaboration 2023). Given the errors, the particle energy lies in the range $E$ = 139−324 EeV (=(1.39−3.24) $\times 10^{20}$ eV).

The values of $R$ for particles with energies of $1.39 \times 10^{20}$ and $3.24 \times 10^{20}$ eV for several source redshifts are given in Table 2. It follows from this table that the errors in the energy of the particle being discussed do not change the above conclusion — its possible sources are at $L \lesssim 5$ Mpc away from us.

Let us turn to the results obtained for secondary silicon particles near the Earth. The energy spectra of the fragments do not coincide, and the intensities of the secondary components at a fixed energy differ by orders of magnitude. The fraction of secondary silicon $P$ is maximal and is $P \sim 0.01$ in a narrow energy range, outside of which it drops by orders of magnitude.

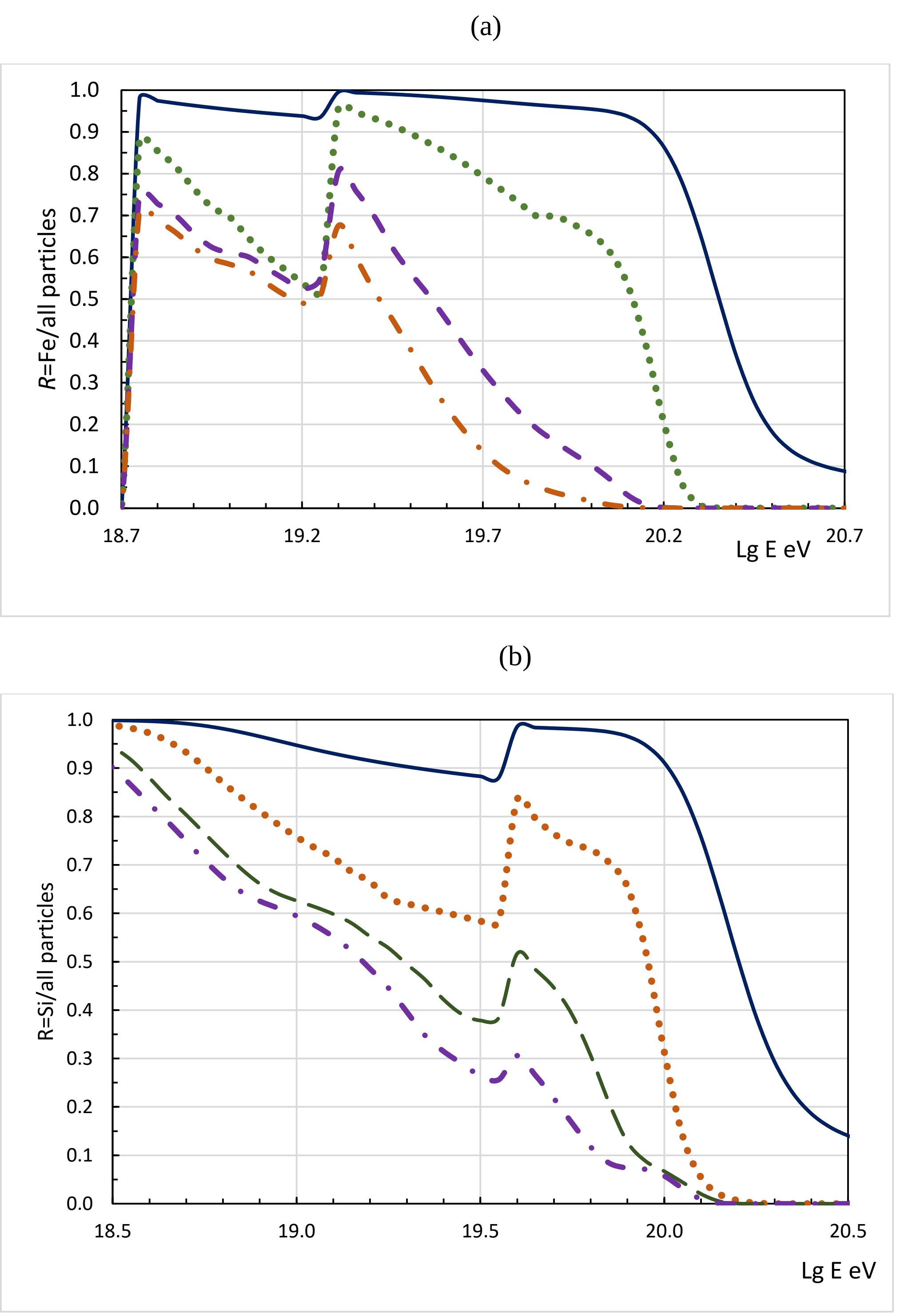

**Fig. 1.** (a) Fraction *R* of iron nuclei that reached the Earth versus energy for various source redshifts *z*: $z = 2.3 \times 10^{-4}$ (solid line), 0.001 (dotted line), 0.005 (dashed line), and 0.009 (dash–dotted line); (b) the same as Fig. 1a, but for silicon nuclei.

On this basis, we sought for the maximum energy of the secondary silicon particles at which $P \geq 0.01$. Next, the energy of the secondary silicon nuclei at which their fraction *P* is maximal depends on the source redshift *z*. Therefore, by specifying different *z*, we can determine how far away the sources of iron nuclei producing secondary silicon particles with a given energy near the Earth are. We obtained the following results from such an analysis. Secondary silicon particles with $E$ = $10^{20}$ eV reach the Earth if the sources of primary CRs are at $L \leq 84$ Mpc away (the redshifts of the sources do not exceed $z \leq 0.019$). For secondary silicon particles with $E \geq 10^{20}$ eV the sources of primary CRs are at $L \leq 55$ Mpc away from the Earth ($z \leq 0.012$).

Let us now consider the particle being discussed with an energy in the range (1.39−3.24) × $10^{20}$ eV by assuming that it is a secondary silicon nucleus produced by an iron nucleus in intergalactic space. The possible sources of such a nucleus are at $L \approx 2-21$ Mpc away from us (the redshifts of the sources are $z \approx 4.5 \times 10^{-4} - 4.5 \times 10^{-3}$). At distances from the source $L$ < 2 Mpc the iron nucleus has no time to produce secondary silicon in reactions (2) and (3).

**Table 1.** The ranges of particle energies $E$ in which $R \gtrsim 0.01$ and the values of $R$ in these ranges for several source redshifts $z$

| z | $2.3 \cdot 10^{-4}$ | 0.001 | 0.005 | 0.009 |
|---|---|---|---|---|
| Fe $E$ eV | (1-6.3) $\cdot 10^{20}$ | (1-2) $\cdot 10^{20}$ | (1-1.41) $\cdot 10^{20}$ | (1-1.12) $\cdot 10^{20}$ |
| Fe $R$ | 0.95-0.08 | 0.65-0.008 | 0.1-0.009 | 0.017-0.008 |
| Si $E$ eV | (1-6.3) $\cdot 10^{20}$ | (1-1.58) $\cdot 10^{20}$ | (1-1.41) $\cdot 10^{20}$ | (1-1.26) $\cdot 10^{20}$ |
| Si $R$ | 0.91-0.12 | 0.31-0.007 | 0.007-0.005 | 0.006-0.009 |

**Table 2.** The values of $R$ for several source redshifts $z$ for iron and silicon nuclei with energies of 1.39 × $10^{20}$ and 3.24 × $10^{20}$ eV

| z | $2.3 \cdot 10^{-4}$ | 0.001 | 0.005 | 0.009 |
|---|---|---|---|---|
| Fe $R(E=1.39 \cdot 10^{20}$eV) | 0.91 | 0.39 | 0.01 | $3 \cdot 10^{-3}$ |
| Si $R(E=1.39 \cdot 10^{20}$ eV) | 0.64 | 0.02 | 0.005 | $3 \cdot 10^{-3}$ |
| Fe $R(E=3.24 \cdot 10^{20}$eV) | 0.18 | 0 | 0 | 0 |
| Si $R(E=3.24 \cdot 10^{20}$ eV) | 0.14 | 0 | 0 | 0 |

## DISCUSSION

In this paper we assume that UHECRs are of extragalactic origin, namely the particles are accelerated in the vicinity of SMBHs. They were detected at the centers of many galaxies, including the Milky Way Galaxy, and at present it is generally accepted that a SMBHs exists in the nucleus of almost every galaxy (Cherepashchuk 2023). (This hypothesis explains the results of our previous studies (Uryson 1996, 1999), where the possible sources of UHECRs were identified with Seyfert nuclei with relatively small jets.)

If this hypothesis about the particle acceleration is correct, then any galactic nucleus with a SMBH within about 21 Mpc can be the source of a heavy nucleus with $E = 2.44 \times 10^{20}$ eV.

Let us find out whether the results obtained depend on the model assumptions about the CR sources.

We assumed that CRs are accelerated in the vicinity of almost every SMBH, but, at the same time, we use a specific type of sources, BL Lac, and their evolution in our calculations. Here, there is no contradiction, and the result obtained does not depend on this assumption. The reason is that the fraction of heavy nuclei near the Earth was calculated for sources with a given redshift $z$, while this fraction is determined only by the propagation of particles in extragalactic space and does not depend on the space density and luminosity of the sources. For the same reason, the shape of the particle injection spectrum adopted in the model does not affect the result.

The characteristics of the extragalactic background light affect the fragmentation of nuclei. In this paper we did not set the goal to find out how $R$ and $P$ change in different background radiation models by choosing a realistic model for each type of radiation.

Magnetic fields in the Galaxy and extragalactic space deflect charged CR particles, and the particle trajectories need to be analyzed to identify the sources of heavy nuclei in UHECRs. The propagation of CRs in galactic and intergalactic magnetic fields is investigated in a number of papers (see, e.g., Dolgikh et al. 2024; Morejon and Kampert 2025; and references therein).

The existing particle detection methods do not allow one to discriminate between iron nuclei ($A = 56$) and their fragments with mass numbers $A = 28-55$. The distances to the sources can seemingly be constrained by analyzing the quantity

$$R1 = (\text{particles with } A = 28-56) \,/\, (\text{all particles with } A = 1-56). \quad (7)$$

It turned out that the shape of the curves $R1(E)$ hardly changes with source redshift, while the value of $R1$ is $\approx 0.8-1$ in the entire range of energies with which the particles with $A = 28-56$ reached the array. Therefore, such an analysis turned out to be noninformative.

## CONCLUSIONS

Primary iron and silicon nuclei with ultra-high energies $E \geq 10^{20}$ eV and $E >\approx 2 \times 10^{20}$ eV can arrive at the array from distances up to 40 Mpc and a distance that does not exceed 5 Mpc, respectively.

Secondary silicon nuclei can be produced in intergalactic space in the interactions of iron particles with background radiations. Secondary silicon nuclei with energies $E \geq 10^{20}$ eV reach the Earth if the sources are $L \leq 55$ Mpc away ($z \leq 0.012$); the sources of particles with $E = 10^{20}$ eV are $L \leq 84$ Mpc away (the redshifts of the sources are $z \leq 0.019$).

In this paper we assumed that the heavy CR particle with an energy (given the errors) $E = (1.39-3.24) \times 10^{20}$ eV recorded by the Telescope Array could be either a primary iron or silicon nucleus or a secondary silicon particle produced by a primary iron nucleus in intergalactic space. The possible sources of this particle are $L \lesssim\approx 5$ Mpc and $\approx 2-21$ Mpc away from us in the former and latter cases, respectively. The fraction of secondary silicon at the energies under consideration is $P \sim 0.01$, and the possibility of recording a secondary nucleus is not ruled out. Since the primary and secondary silicon nuclei are indistinguishable, the constraint on the distance to the possible source is $L \lesssim\approx 21$ Mpc.

In our model the sources of UHECRs are SMBHs in galactic nuclei. According to present views, a SMBH exists in the nucleus of almost every galaxy (Cherepashchuk 2023). Local Supercluster galaxies and Local Group galaxies are then the sources of primary iron nuclei with energies $E > 4 \times 10^{19}$ eV and $E > 10^{20}$ eV, respectively.

Any galactic nucleus with a SMBH within 21 Mpc can probably be the source of a heavy nucleus with $E = 2.44 \times 10^{20}$ eV.

In our model the nonuniformity in the spatial distribution of SMBHs can mainly be responsible for the UHECR anisotropy.

To test the proposed hypothesis about the possible source of the particle with an ultra-high energy of $2.44 \times 10^{20}$ eV recorded by the Telescope Array, it is necessary to calculate the trajectories of UHE iron particles in intergalactic and galactic magnetic fields.

## ACKNOWLEDGMENTS

I thank O. Kalashev for the discussion of the TransportCR code and P. Volchugov who assisted in the calculations. I thank the referee for the discussion and remarks.

## CONFLICT OF INTEREST

As author of this work, I declare that I have no conflicts of interest.

*Translated by V. Astakhov*